\documentclass[aps, twocolumn, showpacs]{revtex4-2}
\begin{document}
\title{New Approach to Unimodular Relativity}
\author{S C Tiwari}
\affiliation{Department of Physics, Institute of Science, Banaras Hindu University, Varanasi 221005, and \\ Institute of Natural Philosophy \\
Varanasi India\\ Email address: $vns\_sctiwari@yahoo.com$ \\}
\begin{abstract}
A thorough study and analysis on the conceptual foundations of unimodular gravity shows that this theory is essentially general relativity disguised as unimodular relativity in the literature. The main reason for this dilemma is accepting the Einsteinian paradigm: general relativistic framework, covariant divergence law for matter energy-momentum tensor, and cosmological constant as an integration constant, but   introducing the artefact of unimodular description absent in Einstein's work. A new approach is proposed in this paper in which pure unimodular relativity is defined in terms of equi-projective geodesics with the fundamental metric tensor having determinant unity and the geometric tensors constructed from them. Modification of covariant divergence law for the matter energy-momentum tensor is shown to have two new consequences. In the conventional unimodular gravity an effective cosmological term comprising of two variable scalar fields, namely, the unimodular geometric ambiguity and unimodular matter energy ambiguity, is proposed. A radical departure on the cosmological constant problem is possible assuming differing evolution of the two scalars: the Einstein equations emerge when the two ambiguities cancel each other. Secondly, in the case of pure unimodular relativity the gravitational field equations are proposed consistent with the unimodular space-time structure.
\end{abstract}
\maketitle
\section{\bf Introduction}

The Standard Model of Cosmology (SMC) is founded on the assumed validity of the Einstein field equations right up to the big-bang singularity, and the belief that the laws of physics tested in the laboratory experiments are applicable not only at the large scale of the Universe but also from the Planck era to the early Universe. In this scenario of SMC and the current stalemate on the Standard Model (SM) in particle physics the nature of space-time structure becomes one of the fundamental problems: is space-time illusion or fundamental reality? We believe that space-time has fundamental physical reality \cite{1} and the geometry of space-time has to be understood afresh. Reviewing the status of physical cosmology \cite{2} Peebles admits that the $\Lambda$CDM picture of the Universe is 'close to reality'; however he also brings out to the fore the anomalies in physical cosmology. 

Dark energy and the cosmological constant problem are undoubtedly the most important issues in SMC. Note that the cosmological constant problem prior to the discovery of the accelerated expansion of the Universe in 1998 was why its value, as determined from the assumption that it was equal to the density of dark energy as estimated by observations, was smaller by about $10^{120}$ orders of magnitude than the Planck energy density, introduced as a 'cut off' to make divergent integrals finite in a quantum mechanical calculation of the vacuum energy. Present observational cosmology shows that about 69 per cent of the energy content of the Universe is in the form of dark energy. Is dark energy in the form of the cosmological constant? For a nice discussion on this question we refer to \cite{2}. One of the promising and widely discussed approaches to the cosmological constant problem is that of the unimodular gravity. In the present paper we make an attempt to seek a unimodular space-time picture independent of the general relativistic space-time. The principle of energy conservation played key role in the foundation of general relativity and Einstein's modifications of the gravitational field equations \cite{3}. For the convenience of reference to Einstein's papers in \cite{3} we use (E1) for his 1916 paper on the foundations, (E2) for the work on cosmology in 1917, and (E3) for the modification in 1919 motivated by the consideration of elementary particles. Unfortunately, a well-defined concept of energy in general relativity has remained unsettled. Note that asymptotic symmetry, for example, spatial infinity and null infinity provide only operational definitions of ADM and Bondi energy-momentum respectively. The construction of the energy-momentum tensor from matter fields on the right hand side of the Einstein field equations makes the geometrization of physics incomplete \cite{4}. Einstein himself considers this aspect unsatisfactory \cite{5,6}. 

The ambiguous role of the energy law in unimodular gravity has led to differing and confusing results on the cosmological constant. The first action integral formulation of unimodular gravity is due to Anderson and Finkelstein \cite{7}.  A modified covariant divergence law for the energy-momentum tensor in connection with unimodular relativity was suggested in \cite{8}. Subsequent arguments \cite{9,10} underline the necessity to resolve the problem related with the meaning of the cosmological constant in unimodular gravity. The problem has to be approached seeking compatibility and consistency of the results derived from the metric tensor, field equations, and action integral. Re-visiting Einstein's work \cite{3} new insights are gained in connection with \cite{7,8,9,10}.  The question of the covariant divergence law for the matter energy-momentum tensor is examined afresh, and it is shown that this law is not indispensable. Relaxing this law and following the modification idea \cite{8} the unimodular ambiguity introduced in \cite{9} is discussed. An important new result emerges that the effective cosmological term has two independent scalar field variables: one of them is proposed to be ralated with what we suggest here as the unimodular geometric ambiguity, and the other one is related with the unimodular ambiguity in the matter energy-momentum tensor suggested in \cite{9}. A recent interesting work on the unimodular gravity in the context of the thermodynamics of space-time \cite{11,12} discusses the role of the nonconformability of quantum fields, however, the considerations on the cosmological constant and covariant divergence law for the energy-momentum tensor are unsatisfactory \cite{13} resulting into the conclusions in conformity with \cite{9}. We argue that the nonconformabilty scalar could be identified with the unimodular ambiguity in matter energy, and could play important dynamical role. 

The paper is organized as follows. In the next section pure unimodular relativity is established. In section III the cosmological constant problem in unimodular gravity is discussed in detail. The conservation or non-conservation of energy in general relativity and modified divergence law in unimodular gravity are discussed in section IV. In section V,  new field equations in pure unimodular relativity are proposed. In the last section concluding remarks and possible implications on SMC and thermodynamics of space-time are presented.

\section{Unimodular relativity}

Physical concepts involved in unimodular relativity could be understood better if we realize that Einstein's preference for assuming the determinant of the metric tensor $g_{\mu\nu}$ to be unity
\begin{equation}
\sqrt{-g} = 1
\end{equation}
is merely for simplification and represents a useful choice of the frame of reference. Einstein himself cautions 'it would be erroneous to believe that this step indicates a partial abandonement of the general postulate of relativity' (E1). Eddington \cite{14} remarks about the assumption (1) that 'it is liable to obscure the deeper significance of the theory, and it is not usually desirable to adopt it'. Vast literature on unimodular gravity suffers from this obscurity. The reason for obscurity may be explained using an interesting example. Consider the famous Schwarzschild solution for the vacuum Einstein equations. To obtain the solution, Schwarzschild proceeded with a flat space-time metric and used the unimodular assumption \cite{15}. Authors \cite{15} consider the use of privileged coordinate system advocated by Einstein rather 'awkward'. They give the example of flat space-time metric in spherical coordinate system that has the determinant $-g =r^4 sin^2\theta$. However, the argument is inappropriate because the Riemann curvature tensor for the Schwarzschild metric does not vanish whereas for flat space-time metric this curvature is zero. Einstein's claim is valid: first a general covariant law is established and then the choice (1) is made to simplify calculations and this in no way violates general relativity (E1). It is however, true that coordinates do matter \cite{15} and may lead to radically different physical outcomes. The most striking example is that of the vacuum solution of the Einstein field equations by Weyl and Levi-Civita for the static axially symmetric case outside cylindrical mass that exhibits non-conservation of energy \cite{16} in contrast to the Schwarzschild solution outside spherically symmetric mass with no violation of energy conservation. The total energy and momentum in general relativity has well defined meaning if there is asymptotic symmetry: at spatial infinity for ADM energy-momentum and at null infinity for Bondi energy-momentum. In the case of isolated spherical mass, asymptotic spatial infinity could be used whereas for a cylindrical mass infinity along z-axis  does not satisfy the condition of asymptotic symmetry resulting into the nonconservation of energy \cite{16}. 

The main reason for misunderstanding on unimodular relativity is to view it solely in the context of the cosmological constant. Let it be emphasized that general relativity has two ingredients: geodesic equation for the description of the motion of a point particle in noninertial frame of pseudo-Riemannian space-time geometry, and the geometric law for gravitation in the form of the Einstein field equations. Logically unimodular relativity should have the counterparts to both independently of general relativity. 

We first adopt the approach based on affine geometry \cite{17} in which the geometry of paths is fundamental. A path is any curve defined by $x^\mu(s)$, where s is a parameter the same for all paths, satisfying the differential equation
\begin{equation}
\frac{d^2x^\mu}{d s^2} + \Gamma^\mu_{\nu\sigma} \frac{dx^\nu}{ds} \frac{dx^\sigma}{ds}=0
\end{equation}
If we make a coordinate transformation $x^\mu \rightarrow x^{\prime \mu}$ then the equivalence of two affine spaces of paths is determined using the transformation of the affine connection $\Gamma^\mu_{\nu\sigma} \rightarrow \Gamma^{\prime \mu}_{\nu\sigma}$.
In a special case for the transformation to the Euclidean geometry the curvature tensor defined by
\begin{equation}
B^\mu_{\nu\sigma\lambda} = \frac{\partial \Gamma^\mu_{\nu\sigma}}{\partial x^\lambda} -  \frac{\partial \Gamma^\mu_{\nu\lambda}}{\partial x^\sigma} + \Gamma^\alpha_{\nu\sigma} \Gamma^\mu_{\alpha\lambda} - \Gamma^\alpha_{\nu\lambda} \Gamma^\mu_{\alpha\sigma}
\end{equation}
vanishes \cite{18}. The affine connection in Eq.(2) and the curvature tensor (3) are defined in the affine manifold independently of any metric structure \cite{17,18}. 

If the affine manifold is endowed with a Riemannian metric structure then the geometry of paths in the Riemannian space-time of general relativity with the metric structure given by the line element
\begin{equation}
ds^2 = g_{\mu\nu} dx^\mu dx^\nu
\end{equation}
reduces to the geodesic line, and the affine connection becomes the Christoffel 3-index symbol of the second kind
\begin{equation}
\frac{\partial g_{\mu\nu}}{\partial x^\lambda} - g_{\mu\alpha} \Gamma^\alpha_{\nu\lambda} - g_{\nu\alpha} \Gamma^\alpha_{\mu\lambda} =0
\end{equation}
One may recognize that Eq.(5) is just the statement that the covariant derivative of the metric tensor vanishes \cite{14}. From the definition (3) one can obtain the Riemann curvature tensor and contracted Ricci tensor for the metric geometry \cite{18}. 

The general coordinate transformations on the affine connection give the following relation, see Eq.(3.7) in \cite{18}
\begin{equation}
\Gamma^{\prime\mu}_{\nu\lambda} =\frac{\partial x^{\prime\mu}}{\partial x^\alpha} (\frac{\partial^2x^\alpha}{\partial x^{\prime\nu} \partial x^{\prime\lambda}} + \Gamma^\alpha_{\beta\gamma} \frac{\partial x^\beta}{\partial x^{\prime\nu}} \frac{\partial x^\gamma}{\partial x^{\prime\lambda}})
\end{equation}
Imposing the condition that the determinant
\begin{equation}
|\frac{\partial x^{\prime\mu}}{\partial x^\alpha}| =1
\end{equation}
an equi-projective geometry of paths is obtained \cite{19} in which the contracted affine connection vanishes
\begin{equation}
\Gamma^\mu_{\mu\nu} =0
\end{equation}

In the metric geometry of general relativity the general coordinate transformation results into the transformed metric tensor
\begin{equation}
g^{\prime\mu\nu} =\frac{\partial x^\alpha}{\partial x^{\prime\mu}} \frac{\partial x^\beta}{\partial x^{\prime\nu}}  g_{\alpha\beta}
\end{equation}
The contracted Christoffel symbol is given by, see expression (35.4) in \cite{14}
\begin{equation}
\Gamma^\sigma_{\mu\sigma} =\frac{\partial (ln \sqrt{-g})}{\partial x^\mu}
\end{equation}
One can see that the condition for equi-projective geometry (7) in the metric geometry leads to the unimodular condition (1) and the vanishing of (10). Therefore, the natural geometry for unimodular relativity is equi-projective geometry. The difference between unimodular and general relativity is not trivial, it has markedly differing observational consequences. In general relativity, $\Gamma^\mu_{\nu\sigma}$ interpreted as generalized force in the geodesic equation plays crucial role to determine the planetary orbits, see sections (38)-(39) in \cite{14}. For example, the nonvanishing $\Gamma^2_{21}$ for the spherically symmetric metric tensor will be zero in unimodular relativity.

It is worth reviewing the important points made in \cite{7} which, unfortunately, were not pursued to completion. Later \cite{9} the authors underlined the difference between a constraint and the condition terming (1) as unimodular condition. However, in the subsequent discussion on the field equations this distinction gets blurred resulting into the standard conclusions on the cosmological constant; we discuss this aspect in the next section. Here we note that \cite{7,9} introduce the fundamental metric tensor for unimodular relativity $f_{\mu\nu}$. It has 9 independent components in 4-dimensional space-time, its determinant has the value $-1$ in all coordinate systems, and it defines the light geometry
\begin{equation}
dl^2 = f_{\mu\nu} dx^\mu dx^\nu =0
\end{equation}
The unimodular conformal metric tensor $f_{\mu\nu}$ is used to define a metric tensor $g_{\mu\nu}$
\begin{equation}
g_{\mu\nu} =  \sqrt{\mu} f_{\mu\nu}
\end{equation}
where the measure field $\mu$ gives a non-dynamical determinant $-g =-\mu^2$.

A cellular structure of space-time proposed in \cite{7} could be given a better perspective. What is the significance of invariant volume element in general relativity? Eddington \cite{14} differentiates between invariant volume element in natural measure that is a physical concept entering into the volume integration, and geometric invariant volume element that counts number of meshes  in the space-time region. Einstein explains nicely how natural or physical volume element is measured by solid rods and clocks for infinitely small region of space-time in which special relativity is assumed to hold. Anderson and Finkelstein \cite{7} consider Riemann's concept of a discrete manifold and light-cone structure that measures the hypervolume element up to a scale factor that is determined relative to a standard unit, say, a quartz crystal. Thus, "Nature does provide a fundamental measure" \cite{7}, and in analogy to quantum cellular structure  of size $\Delta p \Delta q  \approx \hbar$ it may be speculated that unimodular relativity indicates a cellular structure of size $\mu (x) \Delta^4 x$. 

The line element defined by (11) shows that unlike general relativity where one has time-like, light-like and space-like line elements obtained from the metric tensor (4) here the light-like structure is given fundamental importance. Note that in spite of Eddington's criticism of Weyl geometry the special role of gauge-invariant zero length of a vector is an attractive idea. In the particle physics oriented approach to unimodular gravity \cite{20} the utility of massless graviton in the linearized gravity and the use of Wigner's little group for graviton have been shown. In fact, this paper provided great stimulus for unimodular gravity to address the cosmological constant problem in the early years of SMC before the discovery of the accelerating expansion.

In the light of the preceding discussion we make a departure from the conventional approach based on what could be called the Einsteinian paradigm and enunciate the principle of pure unimodular relativity. At this stage we may explain the meaning of Einsteinian paradigm: general relativistic framework, covariant divergence law for matter energy-momentum tensor, and cosmological constant as an integration constant. Einstein did not introduce unimodular gravity and insisting on general relativity arrived at the cosmological constant as an integration constant with clarity. In the Appendix, Einstein's reasoning to arrive at this result, namely, the cosmological constant is an integration constant, is summarized following (E3). However, unimodular gravity discussed in the literature follows this paradigm and ends up in obscurity \cite{7,9,11,12,20}.

{\bf UR Proposition:} The coordinate transformations in 4-dimensional space-time satisfying the condition (7) on the geometry of paths define equi-projective geodesics with the fundamental metric tensor $f_{\mu\nu}$ and the affine connection $\tilde{\Gamma}^\mu_{\nu\sigma}$
restriced by the condition (8) define unimodular relativity. Gravitational field equations must be compatible with the space-time geometry of unimodular relativity.

\section{\bf Cosmological constant}

The historical fact that Einstein introduced the cosmological constant term in the field equations is well known (E2). However, there is a lack of clarity on the physical concepts and motivations related with $\Lambda$ in the work of Einstein. The main points are summarized below.

[1] In the expository article on the final theory of general relativity in 1916 (E1) Einstein explained the utility of the choice (1) in simplifying the calculation of the geometric quantities without diluting the principle of general covariance. An important example given by him is that of the contracted Riemann tensor (Ricci tensor) here denoted by $R_{\mu\nu}$
\begin{equation}
R_{\mu\nu} = A_{\mu\nu} + B_{\mu\nu}
\end{equation}
\begin{equation}
A_{\mu\nu} = \Gamma^\alpha_{\mu\sigma} \Gamma^\sigma_{\alpha\nu} - \frac{\partial \Gamma^\sigma_{\mu\nu}}{\partial x^\sigma}
\end{equation}
\begin{equation}
B_{\mu\nu} =- \Gamma^\alpha_{\mu\nu} \Gamma^\sigma_{\alpha\sigma} + \frac{\partial \Gamma^\sigma_{\mu\sigma}}{\partial x^\nu}
\end{equation}
If $\sqrt{-g} =1$ then using (10) we have
\begin{equation}
B_{\mu\nu} = 0
\end{equation}

[2] In 1917 an attempt to build a model of cosmology led Einstein to modify the field equations (E2). To justify the static solution a universal constant $\Lambda$ was introduced
\begin{equation}
R_{\mu\nu} -\frac{1}{2} g_{\mu\nu} R -\frac{1}{2} g_{\mu\nu} \Lambda = \alpha_G T_{\mu\nu}
\end{equation}
Here $\alpha_G = \frac{8 \pi G}{c^4}$. Einstein uses $\kappa =\frac{8 \pi G}{c^4}$; in \cite{11,12} the whole expression $\frac{8 \pi G}{c^4}$ is retained for the coupling constant, and in \cite{9} the convention is used that $c=1$. As to the $\Lambda$-term the form Eq.(17) is adopted following \cite{9}. To avoid confusion, in the Appendix, we use Einstein's notations for the field equations. 

[3] Does electron have an electromagnetic sub-structure? Since the discovery of electron in 1897, Thomson, Poincare, Lorentz and others were deeply concerned with this question. Einstein asked if the gravitational field could be important in this connection in his 1919 paper (E3). He mentions the efforts of Mie and Weyl in this direction. Once again he modified the field equations. Assuming the traceless energy-momentum tensor $E_{\mu\nu}$ for the electromagnetic field on the right hand side of the field equations the geometric quantity has to be traceless. The proposed equation was
\begin{equation}
R_{\mu\nu} -\frac{1}{4} g_{\mu\nu} R = \alpha_G E_{\mu\nu}
\end{equation}
Arguments involving the stability of the structure of the particle and imposing the energy conservation law led him to derive the new field equations (17) proposed in 1917, but now $\Lambda$ appeared as an integration constant not a universal constant 'peculiar to the fundamental law', see Appendix for details.

The above points clearly establish that the unimodular condition played no role in both the modifications (17) and (18). Strangely, even no mention was made of the decomposition (13) and the vanishing of $B_{\mu\nu}$ for unimodular condition in these modifications. It is also noteworthy that the motivation to introduce the universal constant $\Lambda$ in Eq.(17) is conceptually different than the emergence of $\Lambda$ as an integration constant following the traceless Eq.(18). However, the law of general covariance and the covariant divergence law for the matter energy-momentum tensor are respected in all cases.

Attributing the origin of unimodular relativity or unimodular gravity to Einstein is historically inaccurate and conceptually flawed. In an important paper \cite{7} the action principle for unimodular gravity was given and the meaning of unimodular relativity was clearly explained. However, in the later part of the paper the authors became ambiguous; perhaps for this reason Finkelstein et al \cite{9} incorrectly state that unimodular relativity as an alternative theory of gravity was considered by Einstein in 1919. The conceptual blurring occurs because of the additional assumption of local energy conservation in the form of the vanishing of the covariant divergence of the matter energy-momentum tensor $T_{\mu\nu}$ in the trace-free field equation or in the equation of motion derived from the variational principle modifying the Einstein-Hilbert action incorporating the unimodular condition via Lagrange's method of undetermined multipliers. The action principle for unimodular gravity given in \cite{7} is
\begin{equation}
S = \int [R + \Lambda (\frac{\mu}{\sqrt{- g}} - 1) ] \sqrt{- g} d^4x +S_{matter}
\end{equation}
Here $S_{matter} = \alpha_G \int \sqrt{-g} L_m d^4x$.

Einstein's own thinking on $\Lambda$ is very clear: "from the point of view of logical economy" introduction of $\Lambda$ in the field equations has to be rejected, and had Hubble's expansion been discovered 'at the time of creation of general theory of relativity, the cosmologic member would never have been introduced' \cite{21}. Lemaitre's arguments that $\Lambda$ has theoretical necessity did not convince Einstein who asserted that $\Lambda$ implies 'a considerable renunciation of the logical simplicity of the theory' \cite{5}.

Though relatively less known, the cosmological constant term arises naturally in the unified theory of Weyl \cite{17} transcending the difference between the physics at large (cosmology) and physics at the small (electron). In our view, the intriguing role of the energy conservation \cite{8} and the lack of a complete geometric theory of matter \cite{4,5,6} have allowed a great degree of flexibility in the modifications of the Einstein field equations and the interpretation of the cosmological constant in a vast literature on the dark energy.

A careful study on the unimodular gravity \cite{7,9,20} shows that it is essentially general relativity in the guise of unimodular gravity. Anderson and Finkelstein \cite{7} use $g_{\mu\nu}$ as an auxiliary construct and assume all the geometric tensors constructed from this metric tensor in the action (19), and follow the steps that Einstein took to get the cosmological constant as an integration constant.  The authors in \cite{20} take specific example of the energy-momentum tensor for the scalar field in the action but follow exactly the Einsteinian paradigm. Is there a physical difference between unimodular gravity and general relativity \cite{9}? The authors fail to resolve this question mainly due to their assumption of the covariant divergence law for energy. They introduce an interesting idea of unimodular ambiguity, however the field equations derived from the action (19) are the ones proposed by Einstein, i. e. Eq.(17) where
\begin{equation}
T^{\mu\nu} = \frac{1}{\sqrt{-g}} \frac{\delta \sqrt{-g} L_m}{\delta g_{\mu\nu}}
\end{equation}
The trace of Eq.(17) gives
\begin{equation}
\Lambda = -\frac{1}{2} (R + \alpha_G T)
\end{equation}
Substituting $\Lambda$ in Eq.(17) gives
\begin{equation}
R^{\mu\nu} - \frac{1}{4} g^{\mu\nu} R = \alpha_G ( T^{\mu\nu} -\frac{1}{4} g^{\mu\nu} T)
\end{equation}
Einstein begins with the proposed traceless equations and arrives at Eq.(17), but now the cosmological constant is an integration constant. Van der Biz et al \cite{20} also repeat the same steps as that of \cite{7,9}. Compared to this kind of unimodular gravity Einstein strictly adhered to general relativity even while the traceless equations were given, and for this reason viewed $\Lambda$-term a logical flaw in the gravitational field equations.

Let us now examine the role of covariant divergence law for the energy-momentum tensor \cite{8}. In the development of the gravitational field equations the assumption that
\begin{equation}
T^{\mu\nu}_{~:\nu}=0
\end{equation}
led to the search for a geometric quantity consistent with this assumption, and the Einstein tensor was discovered
\begin{equation}
G^{\mu\nu} = R^{\mu\nu} - \frac{1}{2} g^{\mu\nu} R
\end{equation}
Purely from the geometric view $R^{\mu\nu},~R,~G^{\mu\nu}$ are fundamental tensors constructed from the fundamental metric tensor $g^{\mu\nu}$. A lengthy but straightforward calculation of the covariant divergence of $G^{\mu\nu}$, see Section(52) in \cite{14}, shows that
\begin{equation}
G^{\mu\nu}_{~:\nu} =0
\end{equation}
In the variational formulation assuming the Einstein-Hilbert action
\begin{equation}
S_e =\int R ~\sqrt{-g} d^4x
\end{equation}
and using the Hamiltonian derivative gives
\begin{equation}
\frac{\delta R}{\delta g_{\mu\nu}} = - G^{\mu\nu}
\end{equation}
Eddington makes a general statement for any invariant $K$ giving rise to a symmetrical tensor taking the Hamiltonian derivative with respect to $g_{\mu\nu}$. If we consider the infinitesimal general coordinate transformations then the variational principle for the action $S_e$ directly leads to Eq.(25), i. e. the vanishing of the covariant divergence of the Einstein tensor as a geometric law. Therefore, for the pseudo-Riemannian space-time geometry of general relativity   the law (25) is independent of any consideration on the energy-momentum tensor. The Einstein field equations
\begin{equation}
G^{\mu\nu} = \alpha_G T^{\mu\nu}
\end{equation}
satisfy both (23) and (25). In the modified field equations (17) or the ones derived from the action (19) the natural logical step should be to modify (23) to be consistent with the Bianchi identity (25) that is a characteristic of the Riemannian geometry. Thus following \cite{8}
\begin{equation}
T^{\mu\nu}_{~:\nu} = -\frac{1}{2} \Lambda_{,\nu} g^{\mu\nu}
\end{equation}
Now, the covariant divergence of the traceless equation (22) making use of (21) and (29) gives an identity, and the trace of Eq.(22) too is just $0=0$. There is no cosmological constant. An ingenious proposition to circumvent this argument is that of unimodular ambiguity \cite{9}. In the next section we discuss this issue in detail.

\section{\bf Non-conservation of energy}

Is the covariant divergence law for matter energy-momentum tensor sacrosanct? Could one free unimodular gravity from the burden of a cosmological constant? The first of the two questions has found extensive discussion in the literature and continues to be debated and investigated. The second question is hardly ever discussed. Regarding global energy-momentum conservation in general relativity there is as yet an unsettled interpretation of what constitutes the gravitational energy. In the context of unimodular gravity a modified covariant divergence law for matter energy-momentum tensor was proposed \cite{8}. An interesting consequence is that $\Lambda$ becomes superfluous or a variable field. A serious and thorough examination of the proposed modified law \cite{8} was undertaken by Finkelstein et al \cite{9}. Revisiting the action integral \cite{7} a unimodular ambiguity was added to Eq.(19) by them. The authors \cite{9} concluded that $\Lambda$ was still a constant of integration. Alonso-Serrano and Liska \cite{11} offer an interesting new perspective on the thermodynamics of space-time, and derive the Einstein field equations using local causal diamonds and the concept of entanglement entropy. The authors argue that unimodular gravity rather than general relativity is more natural in the thermodynamical approach. In a recent review \cite{12} a detailed discussion on the conservation of energy-momentum is included; however, following the trodden path on this question the authors reached the same conclusion as the one arrived at in \cite{9}. 

In \cite{10,13} we have questioned this conclusion; however, the possible new physics implied by our comments needs to be delineated in detail. In general relativity it is argued that the covariant divergence law (23) is a consequence of the invariance of the matter action under infinitesimal general coordinate transformations \cite{9,22}. Bak et al \cite{23} gave a detailed discussion on the derivation of the symmetric energy-momentum tensor using Noether's theorem. However, there are at least three arguments that show that the covariant divergence law for the energy-momentum tensor is not indispensable in general relativity.

[A] In the pseudo-Riemannian space-time geometry of general relativity the covariant divergence law for the Einstein tensor (25) has a kind of absolute attribute, and it is independent of the field equations for the metric tensor, i. e.  the vacuum field equations derived from the action (26)
\begin{equation}
G^{\mu\nu} = 0
\end {equation}
On the other hand, to derive Eq.(23) from the variational principle the equations of motion for the physical fields are used to drop the contribution of the variations of the fields under the coordinate transformations \cite{22}. The derivation based on Noether's theorem \cite{23} also has certain unresolved issues being debated in the literature that show that unlike (25) there is no compelling reason for Eq.(23) in general relativity.

[B] Three important arguments put forward in \cite{24} question the assumption (23): extending a physical law that holds in special relativity to general relativity is not unambiguous, the construction of the Lagrangian density in the action for matter fields is open to criticism, and Synge's assumed classical statistical model of the matter in which particles undergo collisions but obey the geodesic equation between the collisions is unjustified.

[C] The separation of gravitational energy from the matter energy is most troublesome. Bondi \cite{16} terms Eq.(23) as a law of non-conservation because the intangible energy of the gravitational field is not accounted for. In Newtonian theory energy is localizable in contrast to the non-localizablity of gravity in general relativity. The example of Weyl-Levi Civita cylindrical metric is discussed in detail and the Newtonian theory and general relativity are compared: in the former only potential energy of the system matters whereas in the later the total energy, matter plus gravitation, is important.  

The above arguments do not prove that covariant divergence law of energy (23) is wrong or invalid; the arguments do show that this law is not as rigid as the Bianchi identity (25), and could be modified. Keeping this perspective in mind let us examine the unimodular ambiguity \cite{9} added to the action (19)
\begin{equation}
S \rightarrow ~ S +\Delta_m S
\end{equation}
\begin{equation}
\Delta_mS = \int \sqrt{-g} ~ \Delta_mL~ d^4x
\end{equation}
This term is supposed to have no observable consequences. The idea is that the action is written in terms of $\sqrt{-g},~ g_{\mu\nu},~R$ and the unimodular condition is also incorporated using $\sqrt{-g}$, therefore, it is not a true unimodular action constructed using $f_{\mu\nu}$. A possible remedy suggested in \cite{9} is to introduce unimodular ambiguity (32). Assuming a simple expression for $\Delta_mL$
\begin{equation}
\Delta_mL = [\frac{\mu}{\sqrt{-g}} -1]~l_m
\end{equation}
the field equations are derived using the variational principle. Here $l_m$ is some function of $g_{\mu\nu}$ and matter variables. Varying $g_{\mu\nu}$ in the action $S+\Delta_mS$ we get
\begin{equation}
R^{\mu\nu} - \frac{1}{2} g^{\mu\nu} R - \frac{1}{2} \Lambda g^{\mu\nu} = \alpha_G (T^{\mu\nu} -\frac{1}{2} l_m g^{\mu\nu})
\end{equation}
Next, a modified energy-momentum tensor is defined
\begin{equation}
T^{\prime \mu\nu} = T^{\mu\nu} -\frac{1}{2} l_m g^{\mu\nu}
\end{equation}
Substituting (35) in (34) the resulting equations are look-a-like modified Einstein equations (17) in which the energy-momentum tensor is $T^{\prime\mu\nu}$. Proceeding as usual adopting Einsteinian paradigm, taking the trace and substituting $\Lambda$ in the field equations the traceless equations are obtained
\begin{equation}
R^{\mu\nu} -\frac{1}{4} R g^{\mu\nu} = \alpha_G (T^{\prime \mu\nu} -\frac{1}{4} T^\prime g^{\mu\nu})
\end{equation}
The authors point out that the re-defined energy-momentum tensor satisfies the modified covariant divergence law given in \cite{8}
\begin{equation}
\alpha_G T^{\prime \mu\nu}_{~:\nu} = -\frac{1}{2} g^{\mu\nu} \Lambda_{,\nu}
\end{equation}
They further assert that $T^{\mu\nu}$ should satisfy (23) in view of the derivation given in \cite{22}, the equation of motion does no depend on $l_m$, and the cosmological constant is a constant of integration.

We criticized this conclusion \cite{10} that it was an assumption not a consequence of the theory. An interesting point in \cite{20} merits attention in this connection. The tensor $T^{\mu\nu}$ is arbitrary up to an additive term if its covariant divergence vanishes (23)
\begin{equation}
T^{\mu\nu} \rightarrow ~ T^{\mu\nu} + C  g^{\mu\nu}
\end{equation}
where $C$ is an arbitrary constant; this constant is not related with the quantum vacuum expectation value of $T^{\mu\nu}$ and can be absorbed in $\Lambda$. Let us consider Einstein field equations (28): in view of the Bianchi identity (25) $G^{\mu\nu}$ is arbitrary up to an additive term $\frac{1}{2} \Lambda g^{\mu\nu}$ if $\Lambda$ is assumed constant, and similarly due to the assumption (23) there exists arbitrariness (38). In the original modification (E2) Einstein postulates $\Lambda$ to be a universal constant. Logically it is possible that instead of absorbing $C$ in $\Lambda$ both may mutually cancel resulting into the unmodified original Einstein equations (28). The problem arises if $\Lambda$ is treated as an integration constant either beginning with the tracefree field equations as Einstein did (E3) or using the action principle that incorporated the unimodular condition through Lagrange's multiplier \cite{7,9,20}. Now, the assumption (23) acquires a fundamental role to introduce at least one arbitrary constant $C$ or $\Lambda$ or $C+\Lambda$ in \cite{7,9,20}.

Does unimodular ambiguity resolve this problem? Consider Eq.(34), taking its trace
\begin{equation}
\Lambda = -\frac{1}{2} (R+\alpha_G T) +\alpha_G l_m
\end{equation}
Recall that the trace of the Einstein field equations (28) determines the scalar curvature
\begin{equation}
-R =\alpha_G T
\end{equation}
In Eq.(39) we define an effective cosmological constant
\begin{equation}
\Lambda^{eff} =\Lambda - \alpha_G l_m
\end{equation}
Using (41) Eq.(34) is re-written as
\begin{equation}
R^{\mu\nu} -\frac{1}{2} g^{\mu\nu} R -\frac{1}{2} \Lambda^{eff} g^{\mu\nu} = \alpha_G T^{\mu\nu}
\end{equation}
tracing this
\begin{equation}
\Lambda^{eff} =-\frac{1}{2} (R+\alpha_G T)
\end{equation}
Eliminating $\Lambda^{eff}$ from Eq.(42) we get the traceless Eq.(22) for $T^{\mu\nu}$. It is easy to verify that $T^{\mu\nu}$ satisfies the modified covariant divergence law \cite{8,10}
\begin{equation}
T^{\mu\nu}_{~:\nu} = -\frac{1}{2} \Lambda^{eff}_{,\nu} g^{\mu\nu}
\end{equation}
Thus, it is established that the intermediate steps from Eq.(18) to Eq.(22) in reference 9 turn out to be superfluous as the role of ambiguity in the energy-momentum tensor $T^{\prime\mu\nu}$ is rendered inconsequential. As shown here a consistent calculation leads to the modified covariant divergence laws (37) and (44) with differing cosmological constants $\Lambda$ and $\Lambda^{eff}$ respectively.

We suggest that the basic idea of unimodular ambiguity introduced by Finkelstein et al \cite{9} hints at new physics. The authors explain the physical motivation behind this idea nicely. The  prescription to construct the extended action depends on "appearances"; it is not really a unimodular action constructed from unimodular variables. Therefore, unlike \cite{9}, logically we expect ambiguity in both geometric part and the matter part of the action, and the most natural and simplest identification for geometric unimodular ambiguity is the cosmological term in Eq.(34). For the sake of clarity Eq.(34) is re-written collecting both ambiguities at one place on the right hand side
\begin{equation}
R^{\mu\nu} - \frac{1}{2} g^{\mu\nu} R = \alpha_G T^{\mu\nu} -\alpha_G \frac{1}{2} l_m g^{\mu\nu}+ \frac{1}{2} \Lambda g^{\mu\nu}
\end{equation}
Comparing Eq.(45) with the Einstein field equations (28) two new possibilities are envisaged. 

NP1: The geometric and matter ambiguities balance each other and the effective cosmological constant vanishes
\begin{equation}
\Lambda^{eff} =0
\end{equation}
The original Einstein field equations acquire new significance: the manifest gravitational field  described by the Einstein equations is determined by the underlying unimodular gravity. 

NP2: The second possibility is that non-zero $\Lambda^{eff}$ occurs as a field variable in the gravitational field equations, however taking the covariant divergence of Eq.(45) we get the modified energy law (43). Note that the $l_m$-term has physical significance as it appears in the equation of motion (45) contrary to the view-point in \cite{9}.  

The remarkable new significance of the unimodular ambiguity seems to be important in the context of the thermodynamical approach for the Einstein field equations as pointed out in \cite{13}. Jacobson \cite{25} puts forward a suggestion based on qualitative arguments that the Einstein equations with an arbitrary cosmological constant can be obtained from the hypothesis of maximum vacuum entanglement in the thermodynamics of space-time. A technical point in this work relates with the inclusion of nonconformal matter fields assumed to be asymptotically conformal at short distances. The variation $\delta <K>$ in the definition of entropy is assumed to have an additional term $\delta X$. The scalar $X$ appears in the equation of state, but using the covariant divergence law for the energy-momentum tensor it is absorbed in the curvature scale $\lambda$ to define the cosmological constant $\Lambda$. In the field equations only $\Lambda$ is present. Instead of making this assumption, i. e. fixing the curvature scale by introducing a space-time constant $\Lambda$ \cite{25}, Alonso-Serrano and Liska \cite{11} determine $\lambda$ tracing the equation of state
\begin{equation}
\lambda = \frac{1}{4} R + \alpha_G (\frac{1}{4} \delta <T> - \delta X)
\end{equation}
and substituting it in the equation of state the unimodular traceless equations are obtained. The authors claim that the thermodynamics of space-time leads to unimodular gravity rather than general relativity. Further, the assertion is made in this paper as well as in a subsequent review \cite{12} that $\delta X$ "which measures the nonconformability of quantum fields" does not affect the gravity. Earlier, we also suggested that unimodular relativity was natural in the thermodynamical approach to space-time \cite{26}: null geodesic congruence and discrete space-time in the thermodynamics of space-time nicely fit with the cellular structure of space-time and the metric $f_{\mu\nu}$ forming the physical basis of unimodular relativity \cite{7}. However, we disagree on relegating the role of $\delta X$ to nondynamical one and the assumption of covariant divergence law for the matter energy-momentum tensor in \cite{11,12}. It is not accidental that the $\delta X$ term has formal resemblence with the $l_m$-term, in fact, the nonconformability is a physically better description for the unimodular ambiguity for the matter energy-momentum suggested in \cite{9} since the unimodular metric $f_{\mu\nu}$ defines light geometry and has conformal symmetry. Therefore, we propose that in the effective cosmological constant  we should have $\frac{l_m}{2} ~\leftrightarrow ~\delta X$.

Finally, it may be speculated that one could view transition of NP2 to NP1 assuming that the changes in unimodular geometric ambiguity and noncoformability are different during their evolution, and the mutual cancellation establishes Einstein field equations at some stage. It seems the thermodynamic approach could be a better description to explore this idea; thermodynamic interpretation for the cosmological constant suggested in \cite{26} would acquire a deeper significance as it becomes zero at some point of evolution. The freedom of having two variable scalar fields could be utilized to address the cosmological constant problem in SMC in a radically new way: competing opposed effects on $\Lambda^{eff}$.

\section{\bf Field equations in unimodular relativity}

Unimodular gravity is tied with general relativity: unimodular constraint or condition refers to the determinant of the metric tensor $g_{\mu\nu}$, the traceless equations proposed by Einstein (E3) comprise of the Ricci tensor $R_{\mu\nu}$ and scalar curvature $R$ constructed from $g_{\mu\nu}$, and the action integral for unimodular gravity also depends on $R, ~g^{\mu\nu}, ~\sqrt{-g}$. The generalized unimodular condition \cite{27} does not depart from this construction. Interestingly, Einstein and Rosen \cite{28} modified the field equations to avoid the singularity arising from the vanishing of the determinant of the metric tensor. The problem is that if at some space-time point the determinant of $g_{\mu\nu}$ vanishes then $g^{\mu\nu}$ becomes infinite and the Ricci tensor, for example, takes the indeterminate form $\frac{0}{0}$.  In spite of discussing the decomposition (13) of $R_{\mu\nu}$ it seems Einstein did not make use of $A_{\mu\nu}$ alone for unimodular gravity. Here an analogue of this tensor in unimodular space-time is used to propose gravitational field equations.

Is it possible to formulate a pure unimodular gravity based on the {\bf UR Proposition}? An outline towards this aim is presented in this section. The geometric objects for unimodular space-time geometry are the fundamental metric tensor $f_{\mu\nu}$, the affine connection $\tilde{\Gamma}^{\mu}_{~\nu\lambda}$ and the tensors constructed from them. By defintion we have
\begin{equation}
f^{\mu\nu} f_{\mu\lambda} = \delta^\nu_\lambda
\end{equation}
\begin{equation}
\tilde{\Gamma}^\mu_{~\nu\lambda} = \frac{1}{2} f^{\mu\sigma} [\frac{\partial f_{\nu\sigma}}{\partial x^\lambda} +\frac{\partial f_{\lambda\sigma}}{\partial x^\nu}-\frac{\partial f_{\nu\lambda}}{\partial x^\sigma}]
\end{equation}
\begin{equation}
\tilde{\Gamma}^{\mu}_{~\nu\mu} =0
\end{equation}
Definition (50) for the equi-projective geometry of paths for pure unimodular relativity could also be obtained from the contraction of expression (49). The contracted affine connection is given by
\begin{equation}
\tilde{\Gamma}^\mu_{~\nu\mu} = \frac{1}{2} f^{\mu\sigma} [\frac{\partial f_{\nu\sigma}}{\partial x^\mu} +\frac{\partial f_{\mu\sigma}}{\partial x^\nu}-\frac{\partial f_{\nu\mu}}{\partial x^\sigma}]
\end{equation}
Since $(\mu, \sigma)$ are dummy suffixes and $f^{\mu\sigma}$ is symmetric, the first and third terms on the right hand side  (RHS) of Eq.(51) will cancel. The determinant of $f_{\mu\nu}$ is -1, therefore, its differential is zero. The calculation using the differential of each $f_{\mu\nu}$ and multiplication with its co-factor will result into $f^{\mu\nu}d f_{\mu\nu} =0$. Thus, the remaining term $\frac{1}{2} f^{\mu\sigma} \frac{\partial f_{\mu\sigma}}{\partial x^\nu}$ in (51) also vanishes and we establish Eq.(50).

The covariant derivatives are defined as usual using the metric tensor $f_{\mu\nu}$ and (49). For example, the covariant derivative of a vector field $A^\mu$ is 
\begin{equation}
A^\mu_{~:\nu} = A^\mu_{~,\nu} - \tilde{\Gamma}^\mu_{~\nu\lambda} A^\lambda
\end{equation}
and that of a second rank tensor field is
\begin{equation}
A^\mu_{\nu~:\sigma} = A^\mu_{\nu~,\sigma}- \tilde{\Gamma}^\alpha_{~\nu\sigma} A^\mu_\alpha + \tilde{\Gamma}^\mu_{~\alpha\sigma} A^\alpha_\nu
\end{equation}
The covariant divergence being a contracted covariant derivative using (50) in the contracted (52) gives
\begin{equation}
A^\mu_{~:\mu} = A^\mu_{~,\mu}
\end{equation}
Contracting (53) and noting that the last term vanishes due to (50) we get
\begin{equation}
A^\mu_{\nu~:\mu} = A^\mu_{\nu~,\mu}- \tilde{\Gamma}^\alpha_{~\nu\mu} A^\mu_\alpha
\end{equation}
Substituting (49) in the second term on the RHS of (55) we get $-\frac{1}{2} [\frac{\partial f_{\nu\beta}}{\partial x^\mu} +\frac{\partial f_{\mu\beta}}{\partial x^\nu}-\frac{\partial f_{\nu\mu}}{\partial x^\beta}] A^{\mu\beta}$. For a symmetric tensor $A^{\mu\beta}$ interchange of suffixes $\mu,\beta$ cancels the first and third terms in the bracket and we finally get
\begin{equation}
A^\mu_{\nu~:\mu} = A^\mu_{\nu~,\mu} +\frac{1}{2} \frac{\partial f^{\alpha\beta}}{\partial x^\nu} A_{\alpha\beta}
\end{equation}

In order to determine the analogue of the Ricci tensor we proceed from the contraction of the curvature tensor (3) first in the indices $(\mu,\nu)$, and then in the indices $(\mu, \lambda)$ and get anti-symmetric tensors
\begin{equation}
S_{\mu\nu} = \frac{\partial \Gamma^\alpha_{~\alpha \mu}}{\partial x^\nu} - \frac{\partial \Gamma^\alpha_{~\alpha \nu}}{\partial x^\mu}
\end{equation}
\begin{equation}
B_{\mu\nu} =  \frac{\partial \Gamma^\alpha_{~\mu \nu}}{\partial x^\alpha} - \frac{\partial \Gamma^\alpha_{~\mu \alpha}}{\partial x^\nu}+\Gamma^\alpha_{~\mu \nu} \Gamma^\beta_{~\alpha \beta} - \Gamma^\alpha_{~\mu \beta} \Gamma^\beta_{~\alpha \nu}
\end{equation}
It can be easily verified that $B_{\mu\nu} - B_{\nu\mu} = S_{\nu\mu}$; thus $B_{\mu\nu}$ is symmetric if $S_{\mu\nu}$ vanishes. It has been proved \cite{18} that in the Riemannian geometry this condition is identically satisfied and $B_{\mu\nu}$ could be identified with the Ricci tensor. Adopting the notation $R_{\mu\nu}$ for the usual Ricci tensor we get $R_{\mu\nu} =-B_{\mu\nu}$ from (58); the minus sign appears because of the difference in the definitions of the curvature tensors in \cite{18} and \cite{14}.

In the equi-projective geometry with metric tensor $f_{\mu\nu}$ for the unimodular space-time the analogue of the Ricci tensor is derived from (58) using (49) and (50) to be
\begin{equation}
\tilde{R}_{\mu\nu} = \tilde{\Gamma}^\alpha_{~\mu\sigma} \tilde{\Gamma}^\sigma_{~\alpha\nu} - \frac{\partial \tilde{\Gamma}^\sigma_{~\mu\nu}}{\partial x^\sigma}
\end{equation}
The scalar curvature is defined as $\tilde{R} = f_{\mu\nu} \tilde{R}^{\mu\nu}$.

The covariant divergence law for the energy-momentum can be relaxed, therefore, the simplest field equations for unimodular gravity are proposed to be
\begin{equation}
\tilde{R}_{\mu\nu} = \alpha_G \tilde{T}_{\mu\nu}
\end{equation}
The energy-momentum tensor has to be symmetric and it will have $f_{\mu\nu}$ not $g_{\mu\nu}$ wherever necessary in order to satisfy the covariance under unimodular space-time coordinate transformations. The proposition on the new field equations is based on the method that Einstein originally used to obtain his field equations. Though energy-momentum tensor for a scalar field $\Phi$ is a trivial example we illustrate the construction of $ \tilde{T}^{(\Phi)}_{\mu\nu}$ for it. In flat space-time geometry with the metric tensor $\eta_{\mu\nu}$ the energy-momentum tensor for massless $\Phi$ is
\begin{equation}
T^{(\Phi)}_{\mu\nu} = \Phi^{:\mu} \Phi^{:\nu} - \frac{1}{2} \eta^{\mu\nu} \Phi^{:\alpha} \Phi_{,\alpha}
\end{equation}
In general relativity the energy-momentum tensor for $\Phi$ is obtained by making the replacement $\eta^{\mu\nu} \rightarrow g^{\mu\nu}$ in Eq.(61). In unimodular relativity it will be obtained using $\eta^{\mu\nu} \rightarrow f^{\mu\nu}$ in Eq.(61)
\begin{equation}
\tilde{T}^{(\Phi)}_{\mu\nu} =\Phi^{:\mu} \Phi^{:\nu} - \frac{1}{2} f^{\mu\nu} \Phi^{:\alpha} \Phi_{,\alpha}
\end{equation}

Although it could be guessed that the proposed field equations (60) and the unimodular geodesic equations differ markedly from Einstein field equations and unimodular gravity based on Einsteinian paradigm, detailed calculations to compare them are essential.  We plan to address this aspect in a later work.

An interesting observation on the role of the determinant of the metric tensor $\sqrt{-g}$ in Einstein's work is made to conclude this section: (i) the assumption (1) retaining general relativity (that has been later interpreted as unimodular condition); (ii) this quantity enters in the invariant volume element, and the volume integral for tensors need tensor densities in the integrand; and (iii) multiplying the tensors in the field equation by the square of the determinant of the metric tensor \cite{28} to tackle the singularities thereby requiring tensor densities in the field equation. In pure unimodular relativity considered here the metric tensor $f_{\mu\nu}$ has fixed determinant (equal to 1), and the tensors and tensor densities are equivalent.

\section{\bf Conclusion}

The main results obtained in the present paper are as follows. The conceptual foundation of unimodular gravity shows that this theory is essentially general relativity disguised as unimodular relativity in the literature. A new approach is proposed for pure unimodular relativity defined in terms of equi-projective geodesics with the fundamental metric tensor having determinant unity and the geometric tensors constructed from them. Plausible arguments are presented to show that the covariant divergence law for the matter energy-momentum tensor could be modified. In the case of conventional unimodular gravity this modification leads to an effective cosmological term comprising of two variable scalar fields: a unimodular geometric ambiguity and unimodular matter energy ambiguity or nonconformability of matter fields. For pure unimodular relativity  new gravitational field equations are proposed consistent with the unimodular space-time structure.

As to the outlook on future work there are two important implications of our results that hold promise for an alternative model of cosmology and in connection with the thermodynamics of space-time. In spite of the precision empirical tests the models beyond SM in particle physics are explored by many physicists \cite{1}. One of the unsatisfactory aspects in SM is that there are nearly twenty adjustable arbitrary parameters in the model. Peebles notes that there exist eight free parameters in SMC, and in a pleasant blending of philosophical, conceptual and empirical aspects related with the foundations of $\Lambda $CDM argues for alternatives to SMC \cite{2}. Among the vast literature on alternatives he points out the model based on Modified Newtonian Dynamics (MOND) \cite{29}. Here, a substantially altered $\Lambda$CDM model for the Cosmology could be envisaged based on the idea of an effective cosmological field (not a constant)  $\Lambda^{eff}$ introduced in the present paper.  It offers a rich structure for the cosmological term. Two scalar fields $\Lambda ,~ l_m \leftrightarrow X$ could present a scenario of the evolution of the Universe from the moment of big-bang to the present in a markedly different form than that of SMC. We suggest three eras of the evolution corresponding to $\Lambda$-dominated, equilibrium of the two scalars, i. e. Einstein field equations based, and $l_m$ or $X$ dominated.

Regarding the thermodynamical approach to space-time, in spite of its attractive features the lack of a microscopic statistical foundation is a serious drawback. Most of the arguments are qualitative and speculative, however, Jacobson being aware of these shortcomings points out that, at least partially, the statistical ingredient is used from the holographic analysis in the recent work \cite{25}. If unimodular gravity is natural for the thermodynamics of space-time \cite{11,12,26} then recalling our suggestion in \cite{26} the cellular structure of space-time \cite{7} may be used for investigating statistical metric of space-time. Following de Broglie's idea that the principle of least action is a particular case of the second law of thermodynamics \cite{4}, it may be further speculated that the discrete cellular space-time structure for unimodular relativity could serve the  purpose for a basis for the entropy law beyond Clausius and entanglement entropies \cite{11,12,25}.

{\bf APPENDIX}

In Einstein's 1919 paper (E3) there are three important steps to examine the question if gravity plays a role in the structure of elementary particles; in fact, essentially a charged particle like an electron. 

{\bf A1:} The Einstein field equations with the energy-momentum tensor of the electromagnetic fields on the RHS are
\begin{equation}
R_{\mu\nu} -\frac{1}{2} g_{\mu\nu} R = -\kappa T_{\mu\nu}
\end{equation}
Here we use Einstein's notations (E3) except that $(G_{\mu\nu}, ~ G ) \rightarrow  (R_{\mu\nu}, ~R) ; \Phi_{\mu\nu} \rightarrow F_{\mu\nu}$. Taking the covariant divergence of Eq.(63), using the Bianchi identity, and the set of the Maxwell equations 
\begin{equation}
\partial_\mu F_{\nu \lambda} + \partial_\lambda F_{\mu \nu} +\partial_\nu F_{ \lambda \mu}=0
\end{equation}
we get
\begin{equation}
F_{\mu\nu} F^{\nu \sigma}_{~:\sigma}=0
\end{equation}
Note that the covariant divergence of the antisymmetrical tensor $F^{\mu\nu}$ is
\begin{equation}
F^{\mu\nu}_{~:\nu} = \frac{1}{\sqrt{-g}}\partial_\nu (\sqrt{-g} F^{\mu\nu})
\end{equation}
Eq.(65) implies that the current-density must vanish everywhere. Not only this, the traceless RHS of Eq.(63) shows that $R=0$ everywhere.
Einstein concludes that based on Eq.(63) a theory of electron is not possible.

{\bf A2:} Instead of Eq.(63) the trace-free field equations are proposed
\begin{equation}
R_{\mu\nu} -\frac{1}{4} g_{\mu\nu} R = -\kappa T_{\mu\nu}
\end{equation}
Evidently the trace of Eq.(67) gives the identity $0=0$. However, the covariant divergence of Eq.(67) leads to an interesting result
\begin{equation}
R_{,\mu} = -4 \kappa ~F_{\mu\nu} J^\nu
\end{equation}
In the space-time regions where the current-density $J^\mu =0$, Eq.(68) shows that the scalar curvature $R$ becomes a constant of integration. Introducing the idea of world-threads such that $J^\mu$ is non-vanishing in the interior of the world-threads, Einstein argues that  $R=R_0$ outside the world-threads where $J^\mu =0$, and the scalar curvature is $R-R_0$ inside the particle. Taking the inner product in Eq.(68) with $J_\mu$, and using the current-density for the particle $ \rho \frac{d x_\mu}{ds}$ Eq.(68) becomes
\begin{equation}
R_{,\mu} \frac{d x_\mu}{d s} =0
\end{equation}
Einstein suggests that $R-R_0$ plays the role of a negative pressure and its fall achieves the stability of the electron.

{\bf A3:} Now, using $R_0$ the trace-free field equations (67) are re-written in the form
\begin{equation}
R_{\mu\nu} -\frac{1}{2} g_{\mu\nu} R  +\frac{1}{4} R_0= -\kappa [T_{\mu\nu} + \frac{1}{4 ~\kappa}~ (R-R_0) g_{\mu\nu}]
\end{equation}
Comparing Eq.(70) with the modified Einstein field equations with $\lambda$
\begin{equation}
R_{\mu\nu} -\frac{1}{2} g_{\mu\nu} R + \lambda g_{\mu\nu}= -\kappa T_{\mu\nu}
\end{equation}
the important result is derived that the cosmological constant $\lambda$ may be identified with $\frac{1}{4} R_0$, and in view of the fact that $R_0$ is an integration constant, the cosmological constant is also an integration constant. Einstein's conclusion in his own words (E3) is worth reproducing: "But, the new formulation has this great advantage that the quantity $\lambda$ appears in the fundamental equations as a constant of integration, and no longer as a universal constant peculiar to the fundamental law." 

{\bf Remarks:} Though Einstein refers to Mie and Weyl no attempt is made by him to place {\bf A2} in the context of  a thorough discussion on the electron theory and the concept of "world-canal" in Section 33 of \cite{17}. Note that in Weyl's unified theory of gravitation and electromagnetism, the scalar curvature $^*R =R+ 6 A^\mu A_\mu -6 A^\mu_{~:\mu}$ assumes importance for the geometric structure of the electron \cite{17}.  A recent discussion could be found in \cite{4}. 

{\bf ACKNOWLEDGMENT}

I thank one of the Editorial Board Members for a painstaking review and the suggestions for the improvement of the presentation of the paper.

\end{document}